\documentclass{article}
\usepackage[dvips]{graphicx}
\usepackage{amssymb}
\usepackage{amsmath}

\begin{document}
\title{Simple models exhibiting spontaneous symmetry breakdown in classical and non-relativistic quantum mechanics}
\author{R. Mu\~{n}oz,$^{1,3,4}$A. Garc\'{i}a-Quiroz,$^{1,5}$,\\ Ernesto L\'{o}pez-Ch\'{a}vez$^{1,6}$ and Encarnaci\'{o}n Salinas-Hern\'{a}ndez.$^{2,7}$\bigskip\\ $1$ Universidad Aut\'{o}noma de la Ciudad de M\'{e}xico,\\Centro Hist\'{o}rico, Fray Servando Teresa de Mier 92, \\Col. Centro, Del. Cuauht\'{e}moc, M\'{e}xico D.F, C.P. 06080\bigskip\\ 
$2$ ESCOM-IPN\\
Av. Juan de Dios B\'{a}tiz s/n esquina Miguel Oth\'{o}n de Mendizabal\\ Unidad Profesional Adolfo L\'{o}pez Mateos\\
 Col. Lindavista, Del. G. A. Madero, M\'{e}xico, D. F, C.P. 07738 \bigskip\\
$3$ corresponding author\bigskip\\$4$ e-mail:rodrigo.munoz@uacm.edu.mx, $5$ e-mail:alberto.garcia@uacm.edu.mx\bigskip\\ $6$ e-mail: elopezc-h@hotmail.com, $7$ e-mail: esalinas@ipn.mx\\
}
\maketitle
\begin{abstract}
First, the properties of a classical model of spontaneous symmetry breakdown are analyzed.  Then, the  \emph{pros} and \emph{cons} of some pedagogical non-relativistic quantum-mechanical models, also used to illustrate spontaneous symmetry breakdown,  are discussed. Finally, a simple quantum-mechanical toy model (a spinor on the line, with a spin-dependent interaction) is presented, that exhibits the spontaneous breaking of an internal symmetry. 
\end{abstract}
\section{Introduction}
Whenever the ground state of a given physical system does not exhibit a symmetry that is present in the fundamental equations of that system, it is said that this symmetry has \emph{spontaneously been broken.}\\  
\emph{The spontaneous breakdown of a symmetry} was first noticed in solid state physics and related fields, where it has played an important role in our understanding of phenomena such as superconductivity and ferromagnetism. For an introduction to symmetry breakdown in that context, one may consult \cite{Aravind} and \cite{Wezel}.\\
In 1960 Y. Nambu \cite{Nambu1} offered the conjecture that some of the approximate symmetries observed in relativistic particle physics could be explained as spontaneously broken exact symmetries.The term \emph{spontaneous breakdown of a symmetry} was introduced the following year by M. Baker and S. L. Glashow \cite{BakerGlashow} (see also \cite{Nambu2}.)\\
An apparently unsurmountable objection to Nambu's conjecture was quickly raised by J. Goldstone, A. Salam and S. Weinberg \cite{Goldstone}. This objection, the Goldstone theorem, states that in every physically acceptable (i. e. covariant) theory, the spontaneous breaking of a symmetry brings with it the presence of unwanted massless particles (the so called Goldstone ghosts, or Goldstone bosons.) \\
This in turn lead to the proposition, by P. W. Anderson \cite{Anderson}, that the coupling of the system with a long-range field (such as the electromagnetic one) could remove the Goldstone ghosts from the theory.\\
Finally, in 1964 P. W. Higgs \cite{Higgs} proposed his celebrated mechanism, by means of which the Goldstone bosons are eliminated by coupling the currents associated with the broken symmetry with a gauge field. From then on, the search for traces of the Higgs mechanism has become one of the main obsessions of experimental and theoretical physicist, as is known by every newspaper reader.\\
Of course, there are many fine points we have left out of this very brief historical account. But at least one thing should come out clear: that \emph{spontaneous symmetry breakdown} plays a central role in contemporary high energy physics, so that it is natural, for anyone studying this field for the first time, to be quickly introduced to this phenomenon,  and  for introductory textbooks, such as \cite{Halzen} or \cite{Nachtmann}, to contain simplified models exhibiting \emph{spontaneous broken symmetries} as a first approximation to the subject.\\
Indeed, \cite{Nachtmann} contains an expositions that starts with a classical model of symmetry breakdown: a point-like classical particle moving on the line, under the sole influence of a potential 
\begin{equation}\label{mexican.peasant.hat}
V(x)=\lambda x^{4}-\mu x^{2}
\end{equation} 
(with $\lambda$ and $\mu$ positive constants) an example of the so called \emph{sombrero} type potentials. There are two different positions of stable equilibrium  (and thus, two different \emph{ground states}) for the particle in this model, as can be deduced from taking a glance at figure 1.\\
 In this example, the potential posses a global spatial-inversion, or $\mathcal{P}$-symmetry, i.e. the symmetry associated with the eveness of the potential about the axis $x=0$
$$V(-x)=V(x)\quad \forall x\in\mathbb{R}$$
As the equilibrium positions (each taken separatedly) do not posses this symmetry, so it is said that the $\mathcal{P}$-symmetry is violated by the (twice-fold degenerate) ground level of potential (\ref{mexican.peasant.hat}).\\
\begin{figure}[t]
\label{msbs1}
\begin{center}
\includegraphics[angle=0, width=\textwidth]{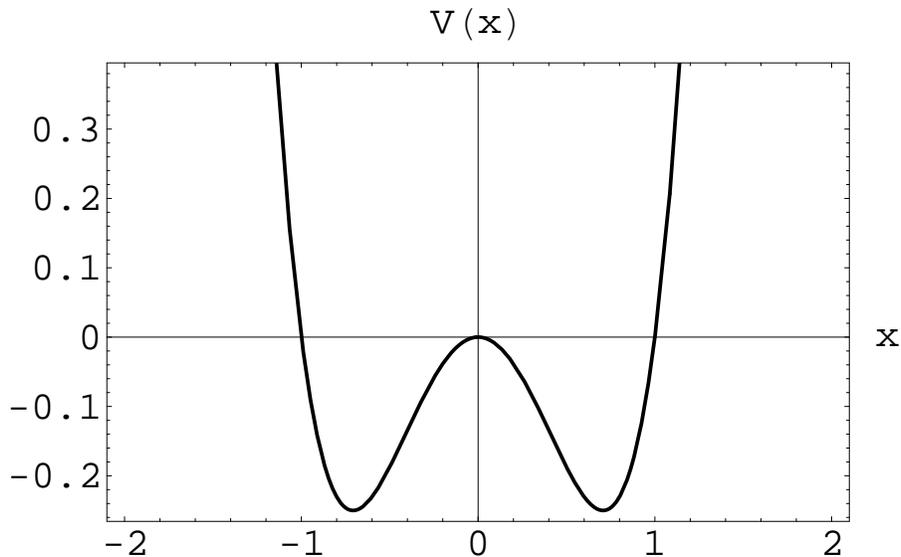}
\end{center}
\caption{The potential $V(x)=x^{4}-x^{2}$ (arbitrary units,) an example of a \emph{sombrero} type potential, with characteristic double minima and reflection symmetry.}
\end{figure} 
Thus, spontaneous symmetry breakdown can be illustrated with a simple model in the realm of classical mechanics. Other, more realistic but less simple, realizations of spontaneous symmetry breakdown in classical systems have been discussed  in \cite{Sivardiere} and \cite{Drugowich}.\\ 
It is then natural to ask, first: If (non-relativistic) quantum-mechanical models can be constructed, that reflect the main features of  spontaneous symmetry breakdown, as they appear in quantum field theories. And, in second term: if this purported non-relativistic quantum models offer any substantial pedagogical advantage over the classical one.\\
Most of this communication is addressed to answering this two questions. We have discovered a previous answer in the well known textbook by Merzbacher \cite{Merzbacher}, which we find to be not completely satisfactory, for reasons that shall be exposed in section 4.\\
This communication has been written with senior university students, and graduate students, in mind. We think it may also be useful for faculty members, and general physicists, interested in obtaining a better grasp of the concept \emph{spontaneous breakdown of a symmetry.} The only requisites for understanding the present communication are a fair level of understanding of classical and non-relativistic quantum mechanics (especially one-dimensional systems) not above the level of the first chapters of a textbook such as \cite{Merzbacher}.\\
\begin{figure}[t]
\label{msbs2_2.1}
\begin{center}
\includegraphics[angle=0, width=\textwidth]{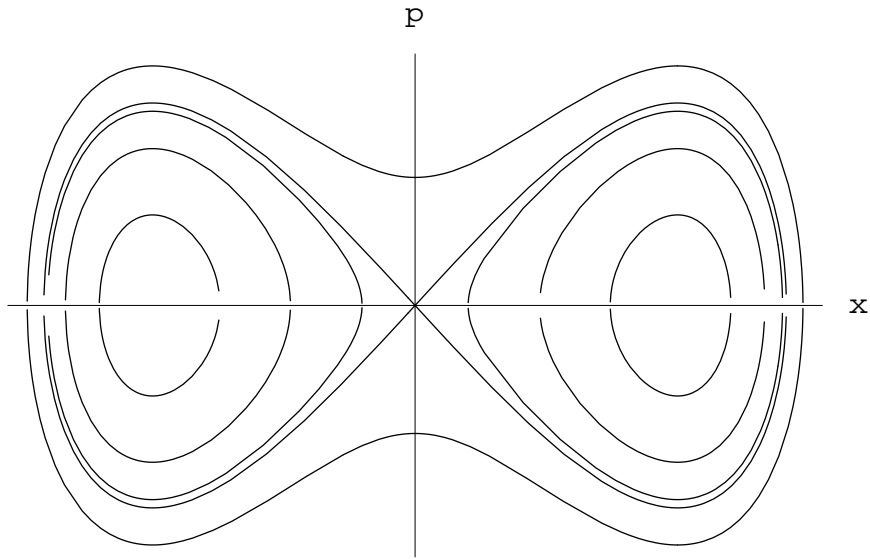}
\end{center}
\caption{The phase portrait of hamiltonian (3.) The separatrix is the curve with condition $E=0$, which crosses the origin, $(x,p)=0$. All trajectories outside the separatrix are symmetric, whereas none of the trajectories inside the separatrix are symmetric.}
\end{figure}  
\section{A comment on the classical model}
Consider the classical model sketched in the previous section, along with its associated hamiltonian:
\begin{equation}
H(x,p)=\frac{p^{2}}{2m}+\lambda x^{4}-\mu x^{2}
\end{equation}
Following Hamilton's formalism, it is not difficult to find that the system obeys equations     
\begin{equation}
\dot{x}=p/m \textbf{  ,  }\dot{p}=2\mu x-4\lambda x^{3}
\end{equation}
and has a phase-space portrait as shown in figure 2. One may very well get the impression, by looking at this phase-space diagram, that the breakdown of symmetry \emph{happens} (so to speak) at separatrix energy $E=0$: above this energy all the phase-space trajectories are symmetric with respect to the $x=0$ axis, while below none of them are. This facts make it cogent that this symmetry breakdown is not necessarily associated with the global $\mathcal{P}$-symmetry of potential $V(x)$, but is rather a \emph{local} phenomenon, that has to do with the properties of $V$ in a neighborhood around a local maximum $x=x_{0}$.\\
Indeed, a potential such as the one shown in figure 3 has, as sombrero-type potentials do, two different positions of stable equilibrium, but in contrast, it has no global spatial-inversion symmetry to be broken. Thus, the very least that can be said is that the degeneracy of the ground state in such classical systems is not necessarily associated with the breakdown of a global $\mathcal{P}$-symmetry. \\
\begin{figure}[t]
\label{msbs3_2.2}
\begin{center}
\includegraphics[angle=0, width=\textwidth]{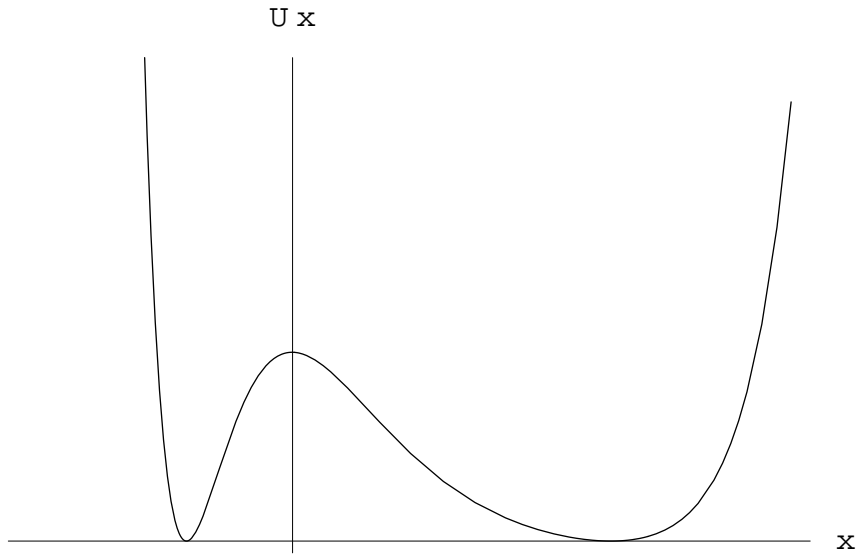}
\end{center}
\caption{The potential $U(x)=\nu\sinh^{2}(\alpha x-3)\sinh^{2}(\frac{1+\alpha x}{20})$ has no reflection symmetry, yet exhibits a twice-fold degenerate \emph{ground level} (two different points of stable equilibrium.)}
\end{figure}  
We ascertain that, on the other hand, there is a breakdown of \emph{local} spatial-inversion symmetry in the phase-space trajectories around the point $(x,p)=(x_{a},0)$  whenever $x_{a}$ is a maximum, for a wide family of analytical potentials.\\
Indeed, let us consider an arbitrary potential $V(x)$, analytical on $\mathbb{R}$. Recalling elementary Calculus, it is not difficult to see that a sufficient condition  for $V(x)$ to have a local maximum at a point $x_{a}$ is that: 
\begin{itemize}
\item[1)] Taylor's expansion around $x_{a}$ takes the form
$$V(x)=V(x_{a})+(x-x_{a})^{2n}\frac{1}{(2n)!}\frac{d^{2n}V}{dx^{2n}}(x_{a})+\mathcal{O}_{2n+1}\quad \textrm{ for some }n\in\mathbb{N}$$
where $\mathcal{O}_{2n+1}$ stands for terms of order equal or higher than $2n+1$, and 
\item[2)]
$$\frac{d^{2n}V}{dx^{2n}}(x_{a})<0$$
\end{itemize}
Thus, $x_{a}$ is a local maximum of $V(x)$ if the behaviour of the system can be approximated by the equation of motion
\begin{equation}\label{repulsiff.f}
m\frac{d^{2}x}{dt^{2}}-(x-x_{a})^{2n-1}\gamma^{2}=0\quad \textrm{ , }\quad \gamma\in\mathbb{R}-\{0\}
\end{equation}
near the point $(x_{a},p_{a})=(x_{a},0)$, with $m$ the mass of the particle. The repulsive force term in (\ref{repulsiff.f}) is derived from the potential
\begin{equation}\label{2.pot}
V_{x_{a}}(x):=V(x_{a})+(x-x_{a})^{2n}\frac{1}{(2n)!}\frac{d^{2n}V}{dx^{2n}}(x_{a})=V(x_{a})-\frac{\gamma^{2}}{2n}(x-x_{a})^{2n}
\end{equation}
which is in all cases \emph{even} about $x_{a}$.\\
At this moment, we can at least ascertain the following: if an analytical potential has a maximum $x_{a}$ complying with conditions (1) and (2,) then there exist a potential (\ref{2.pot}) even about $x=x_{a}$ that is associated with the approximate equation of motion (\ref{repulsiff.f}.)\\
But we can do a little bit better: we can show that in every case when an analytical potential has a maximum as the one described, the approximate equation of motion predicts the existence of two different trajectories for each energy value below the local maximum.\\
 In order to see this more clearly, let us introduce a new variable:
$$x^{\prime}=x-x_{a}$$
in terms of which we have a potential 
$$W(x^{\prime})=V_{x_{a}}(x(x^{\prime}))-V(x_{a})$$
a lagrangian 
$$L(x^{\prime},\dot{x}^{\prime})=\frac{m}{2}\big(\dot{x}^{\prime}\big)^{2}-W(x^{\prime})$$
a momentum
$$p=\frac{\partial L}{\partial x^{\prime}}(x^{\prime},\dot{x}^{\prime})=m\dot{x}^{\prime}$$
and a hamiltonian
$$h(x^{\prime},p)=\frac{p^{2}}{2m}+W(x^{\prime})$$
that give us the equation of motion (\ref{repulsiff.f}) which models the behaviour of the particle in sufficiently small regions around the point $(x^{\prime}=0,p=0)$.  As $h$ has no explicit time dependency, $h=E^{\prime}$ is a constant of motion in each trajectory, and the points of return of a given trajectory must respond to the condition   
$$p=\pm\sqrt{2m\{E^{\prime}-W(x^{\prime})\}}=\pm\sqrt{2m\{E^{\prime}+\frac{\gamma^{2}}{2n}(x^{\prime})^{2n}\}}=0$$
Thus, for energies below $E^{\prime}=0$ there are always two different points of return:
$$x_{\pm}^{\prime}=\pm\sqrt{-\frac{2nE}{\gamma^{2}}}$$ 
one at the left, and one at the right, of the local maximum $x^{\prime}=0.$ This two points belong necessarily to two different, disconnected, trajectories.\\
In so many words we have shown that: for every local maximum, $x_{a}$, of an analytical potential, complying with the previously stated conditions, the $\mathcal{P}$-symmetry of the approximate equation of motion (\ref{repulsiff.f},) that governs the local behaviour of the particle near the point $(x=x_{a},p=0)$, is broken for energies below this local maximum.\\ 
Observe that the symmetry breakdown is better understood, in this case, by considering the phase-space of the system. It is probably fair to say that this illustrates, in the classical realm, Nambu's \emph{dictum:} "...It is always a dynamical question whether a symmetry breaks or not..."\cite{Nambu2}\\  
\section{\emph{Sombrero} type potentials in one-dimensional non-relativistic quantum mechanics}
\begin{figure}[t]
\label{msbs4_3}
\begin{center}
\includegraphics[angle=0, width=\textwidth]{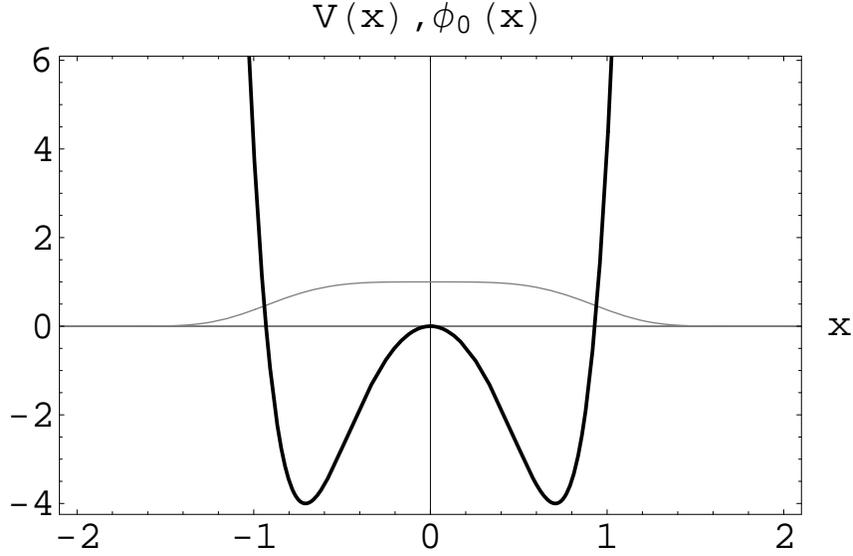}
\end{center}
\caption{The potential of equation (\ref{pottyhat},) shown in black, along with its ground state, shown in gray.} 
\end{figure}
Consider a function $\phi:\mathbb{R}\rightarrow \mathbb{R}$ given by
\begin{equation}
\phi(x)=e^{-ax^{4}}
\end{equation}
for some $a>0$.\\
This function is clearly normalizable, as
\begin{equation}
\int_{-\infty}^{\infty}\vert \phi (x)\vert^{2}dx=\int_{-\infty}^{\infty}e^{-2ax^{4}}dx<\infty
\end{equation}
The operator
\begin{equation}\label{anh}
\hat{a}:=-\imath\frac{d}{dx}-\imath 4ax^{3}
\end{equation}
annihilates $\phi$, \emph{i.e.}
 $$\hat{a} \phi(x)=0$$
so that the explicitly self-adjoint operator
$$\hat{\eta}:=\hat{a}^{\dagger}\hat{a}$$
also annihilates $\phi$.\\
The action of $\hat{\eta}$ on any given function, $\psi (x)$, is easily calculated:
$$\hat{\eta}\psi=(-\imath\frac{d}{dx}+\imath 4ax^{3})(-\imath\frac{d}{dx}-\imath 4ax^{3})\psi=\Big(-\frac{d^{2}}{dx^{2}}+16a^{2}x^{6}-12ax^{2}\Big)\psi$$
Thus, for any value, $m>0$ we can construct a hamiltonian, $\hat{H}$, of the typical Schroedinger form:
\begin{equation}\label{herohat}
\hat{H}:=\frac{\hbar^{2}}{2m}\hat{\eta}=\frac{\hat{p}^{2}}{2m}+V(x)
\end{equation} 
with potential $V(x)$ is given by
\begin{equation}\label{pottyhat}
V(x)=\frac{\hbar^{2}}{2m}\Big(16a^{2}x^{6}-12ax^{2}\Big)
\end{equation}
As $\phi$ is also node-less, we know $\phi$ to be proportional to the ground state eigenfunction of $\hat{H}$. Thus, the ground level energy is zero:
 $$\hat{H}\phi=0$$
Potential (\ref{pottyhat}) is of the \emph{sombrero} type, as is shown in figure 4.\\
In stark contrast with the behaviour of classical systems, the energy eigenvalue of the ground-state (\emph{the most stable stationary state!}) of (\ref{herohat}) corresponds to the energy level of the separatrix  curve of the analogous classical system. Moreover, the probability density of $ \phi$ has its peak, and is centered, in what would be the position of unstable equilibrium in the classical counterpart of potential (\ref{pottyhat}).\\
Yet, the most important point of this example is that there are no stationary states below the even state $\phi$. There is simply no breakdown of symmetry in this case. This is not casual.\\
Indeed, for an analytical  potential $U:\mathbb{R}\rightarrow\mathbb{R}$  with reflection symmetry
$$\mathcal{P}U(x)=U(-x)=U(x) \quad\forall x\in\mathbb{R}$$
stationary states are either even or odd with respect to $x=0$, i.e. for a stationary state $\psi$ of $U$ one can either have
\begin{equation}
\psi_{+}(-x)=\psi_{+}(x) \quad\forall x\in\mathbb{R}
\end{equation}
or 
\begin{equation}
\psi_{-}(-x)=-\psi_{-}(x) \quad\forall x\in\mathbb{R}
\end{equation}
 so that probability density $\vert\psi_{\pm}\vert^{2}$ is in all cases even with respect to $x=0$. In non-relativistic one-dimensional quantum mechanics there are no analytical potentials, \emph{sombrero} type or otherwise, that can exhibit spontaneous $\mathcal{P}$-symmetry breakdown.\\
\section{Spontaneous $\mathcal{P}$-symmetry breakdown in one-dimensional non-relativistic quantum mechanics}
\begin{figure}[t]
\label{cochise}
\begin{center}
\includegraphics[angle=0, width=\textwidth]{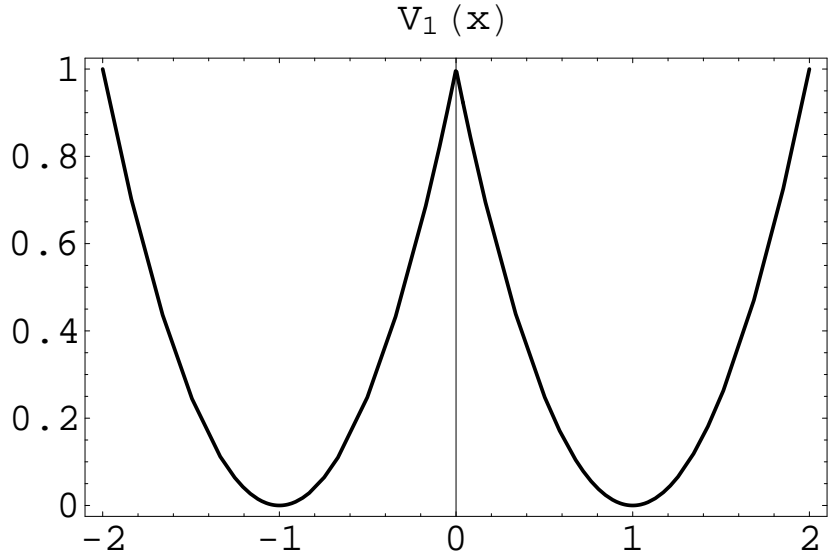}
\end{center}
\caption{An example of the double-oscillator potentials $V_{a}(x)$. Adapted from \cite{Merzbacher}.}
\end{figure}
In the well known textbook by E. Merzbacher \cite{Merzbacher} one can find a model that is claimed to exemplify the spontaneous breaking of symmetry. It starts with a family of double oscillator potentials (see figure 5)
$$V_{a}(x)=m\omega^{2}(\vert x\vert-a)^{2}\quad\textrm { , }\quad a>0$$
and then it is asserted that, as the limit $a\rightarrow \infty$ is approached ``\ldots two degenerate ground state wave functions are concentrated in the separate wells and do not have definite parity . Thus, the reflection symmetry\ldots is said to be hidden or broken spontaneously\ldots"  No proof is provided for this statement.\\
One problem with this model is that there is no potential $V_{\infty}(x)$ that could be represented in graphical form. Indeed, it is not difficult to prove that 
$$\lim_{\alpha\rightarrow\infty}V_{\alpha}(x) =\infty\quad\forall x\in\mathbb{R}$$
It would then seem that in order to accomplish symmetry breaking one would have to have two minima separated by an infinitely wide potential barrier of infinite height.\\
We believe a more graspable example starts with piece-wise-constant potentials of the type
\begin{equation}\label{ucases}
U_{\alpha}(x)=\left \{ \begin{array}{c l}
\infty &\textrm{if } x \leq -a\\
0 &\textrm{if } -b > x > -a\\
\alpha &\textrm{if } b\geq x \geq -b\\
0 &\textrm{if } a> x > b\\
\infty &\textrm{if } x \geq a\\
\end{array}\right .
\end{equation}
(see figure 6) so that, in the limit as $\alpha \rightarrow\infty$, one gets two separated, infinitely deep, wells (see figure 7):
\begin{equation}\label{uinfty}
U_{\infty}(x)=\left \{ \begin{array}{c l}
\infty &\textrm{if } x \leq -a\\
0 &\textrm{if } -b > x > -a\\
\infty &\textrm{if } b\geq x \geq -b\\
0 &\textrm{if } a> x > b\\
\infty &\textrm{if } x \geq a\\
\end{array}\right .
\end{equation}
The advantage being that the solutions of the associated Schroedinger equations can now be worked thoroughly.\\
Each $U_{\alpha}$ ($0<\alpha<\infty$) has a completely discrete spectrum, classified according to parity, with an infinite number of levels above $E=\alpha$. Levels start appearing below the barrier after some threshold value $\alpha_{0}$ is reached in the parameter.\\
Let us focus in the discretization condition below barrier $E=\alpha$. For even states  it reads 
$$E\cot^{2}(a-b)\frac{\sqrt{2mE}}{\hbar}=(\alpha-E)\tanh^{2}b\frac{\sqrt{2m\alpha-E}}{\hbar}$$
\begin{figure}[t]
\label{betty}
\begin{center}
\includegraphics[angle=0, width=\textwidth]{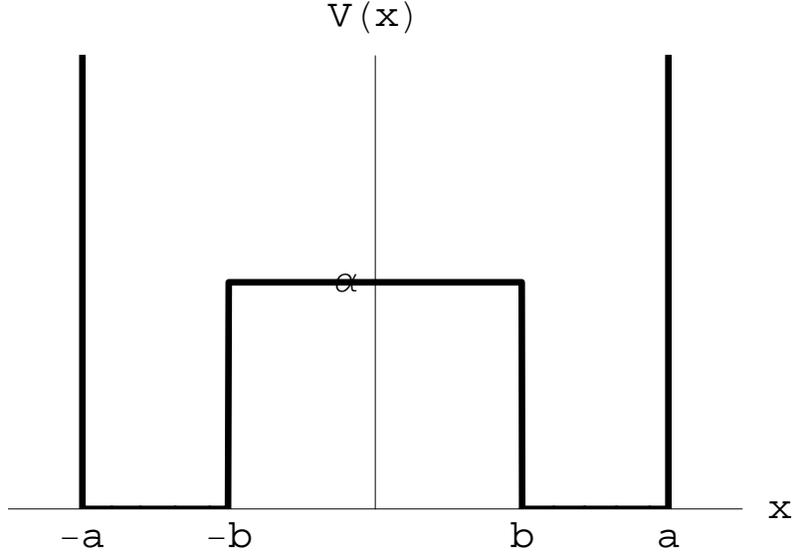}
\end{center}
\caption{An example of the potentials $U_{\alpha}(x)$ of equation (\ref{ucases}).}
\end{figure}
while odd levels below the barrier level have to comply with
$$E\cot^{2}(a-b)\frac{\sqrt{2mE}}{\hbar}=(\alpha-E)\tanh^{-2}b\frac{\sqrt{2m\alpha-E}}{\hbar}$$ 
In the limit $\alpha\rightarrow\infty$ both of this expressions diverge to
$$ E\cot^{2}(a-b)\frac{\sqrt{2mE}}{\hbar}=\infty$$
which only makes sense if
$$E_{n}=\frac{\pi^{2}\hbar^{2} n^{2}}{2m(a-b)^{2}}\quad\textrm{ , } \quad n=1,2,\ldots$$
which, in turn, is the usual discretization condition for a single infinitely deep well of width $a-b$.\\
Thus, for the limit potential $U_{\infty}$ even and odd levels merge, so that the each level is twice-fold degenerate.\\
A pair of perfectly acceptable ``concentrated" eigensolutions, $\psi_{L,n}$ and $\psi_{R,n}$, of the stationary Schroedinger equation:
$$\Big\{-\frac{\hbar^{2}}{2m}\frac{d^{2}}{dx^{2}}+U_{\infty}(x)\Big\}\psi_{j,n}=E_{n}\psi_{j,n} \quad j=L,R$$
are given by
$$\psi_{L,n}(x)=\left\{\begin{array}{cc}
\sqrt{\frac{2}{a-b}}\ \sin \frac{\pi n (x+a)}{a-b}& \textrm{ if } x\in(-a,-b)\\
 & \\
0&\textrm{elsewhere}
\end{array}\right.$$
\begin{figure}[t]
\label{sussy}
\begin{center}
\includegraphics[angle=0, width=\textwidth]{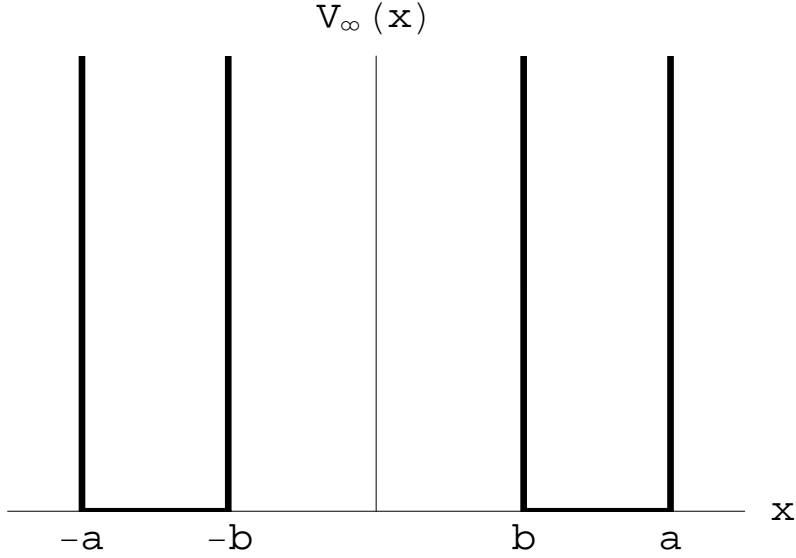}
\end{center}
\caption{The potential $U_{\infty}(x)$ of equation (\ref{uinfty}).}
\end{figure} 
and
$$\psi_{R,n}(x)=\left\{\begin{array}{cc}
\sqrt{\frac{2}{a-b}}\ \sin \frac{\pi n (x-a)}{a-b}& \textrm{ if } x\in(b,a)\\
 & \\
0&\textrm{elsewhere}
\end{array}\right.$$
If the system is in a $\psi_{L,n}$ state, one has absolute certainty that the particle is found at the well standing at the left, and  an analogous assertion can be made about the $\psi_{R,n}$ states and the well at the right. But an equally acceptable pair, spanning the same eigenspace, is
$$\psi_{+,n}=\frac{1}{\sqrt{2}}\Big(\psi_{L,n}+\psi_{R,n}\Big)\quad\textrm{ , }\quad\psi_{-,n}=\frac{1}{\sqrt{2}}\Big(\psi_{L,n}-\psi_{R,n}\Big)$$
the elements of which are not ``concentrated'' in the sense given above.\\
Thus, it is not true that symmetry breaking  \emph{necessarily} begets, in this case, ``concentrated'' states. Rather, the breaking of symmetry brings with it a degeneracy that \emph{allows one to prepare the system in such ``concentrated'' states, among other possible stationary states, including definite-parity ones.}\\
Moreover, the separation between the wells has nothing to do with the breaking of symmetry. In our model, this separation is arbitrary. Changing the distance between the wells, or even taking it to infinity, will not change the behavior of the solutions.\\
This last model is still not completely satisfactory: when symmetry breaking is achieved, at the limit potential, all levels become degenerate, in contrast with the classical model.\\ 
Adding to the strangeness of this last example, the expectation value for the position is, for any of the definite-parity eigenstates $\psi_{\pm,n}$:
$$<\psi_{\pm,n}\vert x\vert  \psi_{\pm,n}>=0$$
which is at the middle of what could be called (tongue in cheek) a \emph{very} classically prohibited region.\\
Let us close this section with yet another quote by Y. Nambu: "A symmetry implies degeneracy. In general there are multiplets...(that) can be distinguished only relative to a weakly coupled external environment which breaks the symmetry." \cite{Nambu2}   
\section{A quantum-mechanical model with spontaneously broken internal symmetry}
Consider a pair of  energy displaced harmonic oscillators:
\begin{eqnarray}\label{twice}
\hat{H}_{+}=-\frac{\hbar}{2m}\frac{d^{2}}{dx^{2}}+\frac{m\omega_{+}^{2}}{2}x^{2}-\frac{\hbar \omega_{+}}{2}\nonumber\\
 \nonumber\\
\hat{H}_{-}=-\frac{\hbar}{2m}\frac{d^{2}}{dx^{2}}+\frac{m\omega_{-}^{2}}{2}x^{2}-\frac{\hbar \omega_{-}}{2}\nonumber\\
\end{eqnarray}
and take inconmensurable fundamental frequencies, \emph{i.e.} take frequencies $\omega_{\pm}>0$ such that the quotient $\omega_{+}/\omega_{-}$ is irrational:
\begin{equation}\label{irrational}
\frac{\omega_{+}}{\omega_{-}}\in\mathbb{R}-\mathbb{Q}
\end{equation}
The matrix arrangement
 $$\mathbb{H}:=
\left(\begin{array}{cc}
\hat{H}_{+}&0\\
0&\hat{H}_{-}\\
\end{array}
\right)$$
which acts on spinors, arrangements of the form
$$\Psi=\left(\begin{array}{c}
\psi_{+}\\
\psi_{-}\\
\end{array}\right)\qquad\textrm{ , }$$
has eigenvalues and eigenspinors given by the equations
\begin{equation}
\mathbb{H}\Psi_{\pm,n}=\hbar n\omega_{\pm}\Psi_{\pm,}
\end{equation}
where 
$$\Psi_{+,n}=\left(\begin{array}{c}
\psi_{+,n}\\
0\\
\end{array}\right)\quad\textrm{ , }\quad 
\Psi_{-,n}=\left(\begin{array}{c}
0\\
\psi_{-,n}\\
\end{array}\right)
$$
the $\psi_{\pm,n}$ ($n\in\mathbb{N}\bigcup \{0\}$) standing for  the well known eigenfunctions of hamiltonian $H_{\pm}$\\
The so called Pauli matrix
$$\sigma_{3}=\left(\begin{array}{cc}
1&0\\
0&-1\\
\end{array}\right)$$
which is self-adjoint and unitary, i.e.
$$u^{\dagger} u=u^{2}=\mathbb{I}$$
($\mathbb{I}$ standing for the $2\times 2$ identity matrix) is an internal symmetry of the system:
$$[\sigma_{3},\mathbb{H}]=0$$
and eigenspinors are classified according with this symmetry:
$$\sigma_{3}\Psi_{\pm,n}=\pm\Psi_{\pm,n}$$
Condition (\ref{irrational}) guarantees that all the excited levels in the spectrum of $\mathbb{H}$ are not degenerate, yet the ground state level is:
$$\mathbb{H}\Psi_{\pm,0}=0$$
Thus, in this model symmetry is broken only at ground state level.\\
It is interesting to note that $\sigma_{3}$, along with the identity matrix, furnishes a representation of the group of second order, just as $\mathcal{P}$ and the identity operator do:
$$\sigma_{3}^{2}=\mathbb{I}\quad\textrm{ just as }\quad \mathcal{P}^{2}=\mathbb{I}$$
This example may be given a little more \emph{physical} (in contrast with purely mathematical) appearance, by writing $\mathbb{H}$ in the form
$$\mathbb{H}=\hat{H}_{0}\mathbb{I}+u(x)\sigma_{3}$$
where $\hat{H}_{0}$ is just an energy displaced harmonic oscillator
$$\hat{H}_{0}=-\frac{\hbar}{2m}\frac{d^{2}}{dx{2}}+\frac{m\omega_{0}^{2}}{2}x^{2}-\epsilon_{0} $$
and $u\sigma_{3}$ is a \emph{position-dependent spinorial interaction}, with position-dependence given by:
$$u(x)= \frac{m \omega_{\Delta}^{2}}{2}x^{2}-\epsilon_{\Delta}$$
In order to do this, one just needs to define:
$$ \begin{array}{cc}
\omega_{0}:=\sqrt{\frac{\omega_{+}^{2}+\omega_{-}^{2}}{2}}\quad\textrm{ , }& \epsilon_{0}:=\frac{\hbar (\omega_{+}+\omega_{-})}{2}\\
 & \\
\omega_{\Delta}:=\sqrt{\frac{\omega_{+}^{2}-\omega_{-}^{2}}{2}}\quad\textrm{ , }&\epsilon_{\Delta}:=\frac{\hbar (\omega_{+}-\omega_{-})}{2}\\
\end{array}$$
Finally, a word of caution is probably in order, more for the seasoned reader than for the beginner: although the toy model discussed in this section may remind of (and is partially inspired in) the quantum-mechanical super-symmetric models presented by Witten in \cite{W}, our system is clearly \emph{not} super-symmetric. The telling difference being that its energy levels are not degenerate above ground-state level, while the excited levels of quantum-mechanical super-symmetric models are \emph{always} degenerate, not withstanding if super-symmetry is broken or not.\\
\section{Conclusions}
In the preceding pages we have shown:
\begin{itemize}
\item[-]the local character of the $\mathcal{P}$-symmetry broken in one-dimensional classical models
\item[-]the impossibility of spontaneous $\mathcal{P}$-symmetry breaking for one-dimensional analytical quantum-mechanical potentials
\item[-]the deficiencies of some of the previous models
\item[-]the viability of simple quantum-mechanical models with spontaneously broken internal symmetries
\end{itemize}

\section{Acknowledgments}
The support of SNI-CONACYT (Mexico) is duly acknowledged. 

\end{document}